\documentclass[         %
aps,                    
prd,                    
showpacs,               
superscriptaddress,    
nofootinbib,            
onecolumn,             %
showkeys,               %
preprintnumbers,        %
floatfix               
]{revtex4}               
\usepackage{graphicx,amsmath}
\usepackage{caption}
\usepackage{subcaption}
\usepackage{enumerate}
\usepackage{natbib}
\usepackage{xcolor}
\usepackage{float}
\usepackage[section]{placeins}

\begin{document}
\title{Symmetric and Standard Matter-Neutrino Resonances Above Merging Compact Objects}
\author{A.~Malkus}
\email{acmalkus@ncsu.edu}
\affiliation{Department of Physics, North Carolina State University, Raleigh, NC 27695 USA.}
\author{G.~C.~McLaughlin} 
\email{gcmclaug@ncsu.edu}
\affiliation{Department of Physics, North Carolina State University, Raleigh, NC 27695 USA.}
\author{R.~Surman}
\email{rsurman@nd.edu}
\affiliation{Department of Physics, University of Notre Dame, Notre Dame, IN 46656 USA.}
\date{\today}
\begin{abstract}
Matter-neutrino resonances (MNR) can occur in environments where the 
flux of electron antineutrinos is greater than the flux of electron neutrinos. 
These resonances may result in dramatic neutrino flavor transformation.  Compact
object merger disks are an example of an environment where electron antineutrinos
outnumber neutrinos. We study MNR resonances in several such disk configurations and 
find two qualitatively different types of matter-neutrino
resonances: a standard MNR and a symmetric MNR.  We examine the transformation that
occurs in each type of resonance and explore the consequences for nucleosynthesis.
\end{abstract}
\medskip
\maketitle
\section{Introduction} 

Neutrinos shape the physical phenomena surrounding compact object mergers, from the dynamics 
of the disk or hypermassive-neutron star itself \cite{Foucart:2014nda,Perego:2014fma, 
Deaton:2013sla,Palenzuela:2015dqa}, to the energetic jets, e.g. \cite{Nagakura:2014hza} that 
may from them. Neutrinos also play an important role in the nucleosynthesis that takes place 
in and around disks \cite{Surman:2003qt,Roberts:2010wh,Wanajo:2014wha,Goriely:2015fqa}. For 
example, the wind outflows \cite{Wanajo:2011vy,Dessart:2008zd} above disks can be home to 
nucleosynthesis, including perhaps the r-process, depending on neutrino flavor composition 
\cite{Surman:2005kf,Surman:2008qf,Caballero:2009ww,Perego:2014fma,Bauswein:2014vfa,Just:2014fka,Goriely:2015fqa}. 
The neutrino flavor composition above the neutrino trapping surface depends not only on 
thermodynamics in the trapped regions, but also the oscillation of neutrinos as they leave the 
disk. The high neutrino density coupled with high matter density provide an environment where 
several kinds of oscillation may take place. Neutrinos emitted from mergers can undergo the 
same types of transformations that neutrinos from supernovae do \cite{Dasgupta:2008cu}, as 
well as oscillations not previously seen elsewhere (except for in collapsars 
\cite{Malkus:2012ts}), called Matter-Neutrino Resonance (MNR) transitions 
\cite{Malkus:2014iqa}. The MNR takes place when the matter potential and the neutrino self 
interaction potential are the same size and have opposite signs.

Much of the previous work on neutrino transformation in high neutrino density environments has 
considered matter and self interaction potentials of the same sign.  For example, for the case 
of core collapse supernovae neutrinos, it has been pointed out that several types of 
transformations can occur in such systems which have a slight electron neutrino excess, start 
at high neutrino density and end at low neutrino density; for recent work see e.g. 
\cite{Strack:2005ux,Pehlivan:2011hp,Cherry:2012zw,Volpe:2013uxl,Vlasenko:2013fja,Duan:2014gfa,Pllumbi:2014saa,Mirizzi:2015fva}. 
At very high densities of neutrinos, synchronized neutrino oscillations, e.g. 
\cite{Duan:2005cp}, change neutrino flavor on a small scale.  At very low densities of 
neutrinos, the neutrino self interaction potential is unimportant and Mikheyev Smirnov 
Wolfenstein (MSW) oscillations can take place when the scale of the matter potential is the 
same as the vacuum scale. In between these extreme regimes of synchronized and MSW 
oscillations, large scale flavor transformation can take place when the neutrino self 
interaction potential approaches the vacuum oscillation scale 
\cite{Hannestad:2006nj,Duan:2007mv} and the neutrinos and antineutrinos enter the transition 
region nearly in flavor eigenstates, e.g. 
\cite{Duan:2007mv,Hannestad:2006nj,Dasgupta:2008cd,Duan:2007sh,Balantekin:2006tg,EstebanPretel:2008ni,Gava:2009pj,Duan:2010bf,Duan:2010af}. 
During the transition, the neutrinos and antineutrinos are said to be ``locked'' as their 
survival probabilities mirror each other and the phenomenon is referred to as ``bipolar'' or 
``nutation'' oscillation.

Unique to settings where the neutrino interaction potential and the matter potential have 
opposite sign, another oscillation phenomenon can be possible 
\cite{Malkus:2012ts,Malkus:2014iqa}. In \cite{Malkus:2012ts} it was shown that collapsar-type 
disks may be home to MNRs and that the MNR may result in flavor transformation that alters 
nucleosynthesis. The understanding of the MNR was expanded in \cite{Malkus:2014iqa}, where the 
standard MNR was explored in detail. A MNR begins when the scale of the matter potential and 
the neutrino self interaction potential are the same and can cancel. A MNR transition can 
cause a dramatic change in the flavor of neutrinos. During the transformation, the neutrino 
self interaction potential matches the size of the matter potential over an extended period of 
time. The transition continues as long as the neutrinos can change flavor in such a way that 
the potential matching is possible. Once the neutrinos can no longer keep up with the matter 
potential, then the transformation ceases.

Whether a MNR region occurs and whether it results in flavor transformation depends upon the 
configuration of the emission surfaces. We examine two qualitatively different types of self 
interaction potentials that can arise from mergers. We point out that while there exist two 
different self interaction potentials, which result in distinct outcomes, the oscillation 
phenomena are both well described as MNR transitions. Our study is presented in this paper as 
follows: First we discuss representative disk configurations in section 
\ref{sec:describeMergers}, then we discuss oscillation calculations in section 
\ref{sec:calculations}. We present neutrino oscillation calculation results in section 
\ref{sec:results} and the results of nucleosynthesis calculations in section 
\ref{sec:nucleosynthesis}. We then conclude in section \ref{sec:conclusion}.

\section{Accretion Disk Configurations}\label{sec:describeMergers}
Similar to core collapse supernovae, the mergers of compact objects like two neutron stars or 
a black hole and a neutron star release vast amounts of energy in the form of neutrinos and 
antineutrinos. However, unlike supernovae, these systems begin with a composition of almost 
entirely neutrons.  After the collision, the material is heated and the electron fraction 
increases.  As a consequence, the resulting emission of antineutrinos is greater than the 
emission of neutrinos. Electron neutrinos and antineutrinos are thought to dominate the flux 
\cite{Foucart:2014nda,Perego:2014fma,Deaton:2013sla,Oechslin:2006uk,Dessart:2008zd} and other 
flavors may not be trapped at all \cite{janka}. Simulations of compact object mergers show 
small amounts of mu and tau neutrinos with a luminosity of up to $1.5\times 10^{52}$ erg/s 
\cite{Foucart:2014nda}. This luminosity is about $1/10^{\text{th}}$ of the luminosity of 
electron antineutrinos.

Guided by these results, we generate a set of representative disk models and calculate the 
neutrino oscillation pattern above them. The disk models used in the following calculations 
are inspired by those in \cite{Setiawan:2005ah,Ruffert:2001gf,Janka:1999qu,Surman:2008qf}, 
for a merger of a $2.5 M_\odot$ black hole and a $1.6 M_\odot$ neutron star that forms a 
black hole with mass $3.85 M_\odot$ and spin parameter $a=0.6$. For ease of calculation, the 
disks are taken to be geometrically thin.  We use two general types of models: one with all 
three types of neutrino emission coming from a disk with a single maximum radius, and a second 
where each the three types comes from a unique disk with its own radius.

Each of the three kinds of neutrinos (electron neutrinos, electron antineutrinos, and all 
other flavors of neutrinos and antineutrinos) are taken to have different temperatures.  These 
temperatures and radii are listed in Table \ref{tab:singleDiskParameters} for our single 
radius model and Table \ref{tab:multiDiskParameters} for our multiple radius model. We further 
vary the amount of $\nu_\mu$, $\nu_\tau$, $\bar{\nu}_\mu$, and $\bar{\nu}_\tau$ within each 
type of model.
\begin{table}
\begin{tabular}{|l |l |l|}
  \hline
  $a$                 & $T_a$  (MeV)& $R_a$ (cm) \\
  \hline
  $\nu_e$               & 6.4          & $4.5\times10^6$\\ 
  \hline
  $\nu_{\mu,\tau}$       & 7.4          & $4.5\times10^6$\\ 
  \hline
  $\bar{\nu}_e$         & 7.1          & $4.5\times10^6$\\ 
  \hline
  $\bar{\nu}_{\mu,\tau}$ & 7.4          & $4.5\times10^6$\\ 
 \hline
\end{tabular}
\caption{\textbf{Single Radius Model}: Parameters for neutrino emission from a single surface.
Fluxes are taken to be thermal Fermi Dirac fluxes for electron neutrino and electron antineutrinos. 
For mu and tau flavor neutrinos and antineutrinos, the flux is taken to be the thermal flux rescaled by an overall parameter specified in each example. 
}\label{tab:singleDiskParameters}
\begin{tabular}{|l |l |l|}
  \hline
  $a$                 & $T_a$ (MeV)& $R_a$ (cm)\\
  \hline
  $\nu_e$               & 5.9          & $5.2\times10^6$\\ 
  \hline
  $\nu_{\mu,\tau}$       & 9.9          & variable\\ 
  \hline
  $\bar{\nu}_e$         & 7.8          & $3.9\times10^6$\\ 
  \hline
  $\bar{\nu}_{\mu,\tau}$ & 9.9          & variable\\ 
  \hline
\end{tabular}
\caption{\textbf{Multiple Radius Model}: Parameters for neutrino emission from several surfaces.
  Fluxes are taken to be thermal Fermi Dirac fluxes for electron neutrino and electron antineutrinos. 
  For mu and tau flavor neutrinos the fluxes are taken to be the thermal fluxes as well, but the radius of the emission is taken to be different as specified in each example.
}
\label{tab:multiDiskParameters}
\end{table}

\begin{figure}
  \includegraphics[angle=270,width=0.5\textwidth]{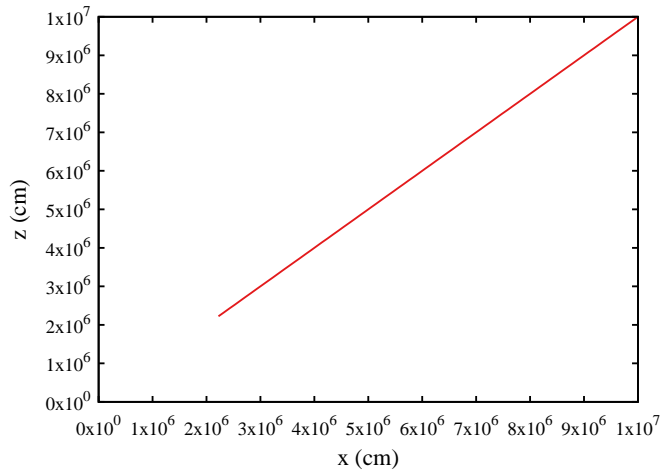}
  \caption{Trajectory of the test neutrino, $\vec{r}=(x,y=0,z)$.}
  \label{fig:trajectory}
\end{figure}

We follow a test neutrino as it leaves each disk model at $45^\circ$ from the plane of the disk. 
The test neutrino starts above the disk and follows a radial trajectory outward, which we show 
in Fig. \ref{fig:trajectory}. The trajectory follows the mass outflow as it leaves from the 
disk. We use a parameterized outflow velocity, $u$, as in \cite{Surman:2004sy}. The velocity 
depends on the acceleration of the material, $\beta$, and the eventual velocity of the material 
at infinity, $v_\infty$,
\begin{equation}
  \left| u\right| = v_\infty\left(1-\frac{R_{inner}}{R}\right)^\beta,
\end{equation}
where $R$ is the distance from the center of the disk and $R_{inner}$ is the initial position 
of the material, which we take to be $R_{inner}=2.0\times 10^6$ cm. As the neutrino travels 
along the trajectory, the neutrino will feel potentials based on its position.  The matter 
potential, due to coherent forward scattering with electrons and positrons, is computed from 
the number density based on the outflow model, which assumes a constant mass outflow rate. The 
outflow model, with $s/k=50$, $\beta=2.0$ and $v_\infty=0.1c$, yields net electron number 
density, $N_e(t)$, where $t$ parameterizes the position along the trajectory. The position 
$t=0$ is the start of the trajectory.

\section{Calculations}\label{sec:calculations}
We calculate the flavor transformation of neutrinos and antineutrinos as they travel along the trajectory which we show in Fig. \ref{fig:trajectory}.
The evolution of neutrinos and antineutrinos is computed through the $S$ matrices, which in the flavor basis are governed by,
\begin{equation}
  i\frac{d}{dt}S(E) = \left(H_V(E) + H_e(t)+H_{\nu\nu}(t)\right)S(E).\label{eq:evolution}
\end{equation}
The vacuum Hamiltonian is given in the flavor basis by,
\begin{equation}
 H_V(E)= U_{23}(\theta_{23})U_{13}(\theta_{13})U_{12}(\theta_{12})\left(
  \begin{array}{ccc}
    -\Delta_{21}(E) & 0 & 0\\
    0 & \Delta_{21}(E) & 0\\
    0 & 0 & \left(\Delta_{31}(E) + \Delta_{32}(E)\right)
  \end{array}
\right)U^\dagger_{12}(\theta_{12})U^\dagger_{13}(\theta_{13})U^\dagger_{23}(\theta_{23}),
\end{equation}
where $\Delta_{ij}(E)=(m_i^2-m_j^2)/(4E)$ and the $U_{ij}(\theta_{ij})$s are the unitary matrices that take the Hamiltonian from the mass basis where $H_V(E)$ is diagonal to the flavor basis.
The matter potential influencing neutrino oscillations is $V_e(t)=\sqrt{2}G_FN_e(t)$, where $G_F$ is the Fermi constant.
The matter potential results in Hamiltonian contribution, $H_e(t)$, 
\begin{equation}
  H_e(t)=\left(
  \begin{array}{ccc}
    V_e(t) & 0 & 0\\
    0&0&0\\
    0&0&0
  \end{array}
  \right).\label{eq:defineHe}
\end{equation}
Neutrinos also feel a potential from interacting with other neutrinos often called the self interaction potential. This potential is computed similarly, but because the neutrinos are not isotropically distributed above the disk, the test neutrino will feel a non trivial geometric effect from the other neutrinos.
The resulting Hamiltonian contribution is $H_{\nu\nu}(t)$,
\begin{equation}
  H_{\nu\nu}(t)=\int_0^\infty\left(S(t,E)\rho(t,E)S^\dagger(t,E)-\bar{S}^*(t,E)\bar{\rho}(t,E)\bar{S}^T(t,E)\right)\,dE. \label{eq:defineHnunu}
\end{equation}
Initially $S$ is the identity matrix, but evolves according to Eq. \ref{eq:evolution} as the neutrinos oscillate.
The matrices, $\rho$ and $\bar{\rho}$ take the form,
\begin{equation}
  \begin{aligned}
    \rho(t,E)=&\sqrt{2}G_F\left(
    \begin{array}{ccc}
      \phi_{\nu_e}(E)C_{\nu_e}(t) & 0 & 0 \\
      0&\phi_{\nu_\mu}(E)C_{\nu_\mu}(t) &0\\
      0&0& \phi_{\nu_\tau}(E)C_{\nu_\tau}(t) 
    \end{array}
    \right)\\
    \bar{\rho}(t,E)=&\sqrt{2}G_F\left(
    \begin{array}{ccc}
      \phi_{\bar{\nu}_e}(E)C_{\bar{\nu}_e}(t) & 0 & 0 \\
      0&\phi_{\bar{\nu}_\mu}(E)C_{\bar{\nu}_\mu}(t) &0\\
      0&0& \phi_{\bar{\nu}_\tau}(E)C_{\bar{\nu}_\tau}(t) 
    \end{array}
    \right),
  \end{aligned}
\end{equation}
where $C_{a}(t)$ is the geometric contribution, and the $\phi_a(E)$ is the flux for the $a=\nu_e,\nu_\mu,\nu_\tau,\bar{\nu}_e,\bar{\nu}_\mu,\bar{\nu}_\tau$ disk.
The geometric contribution was derived in \cite{Malkus:2012ts} for a disk in general.  Since we are calculating trajectories like those in Fig. \ref{fig:trajectory}, which make a $45^\circ$ angle with the plane of the disk, we can reduce this general expression to
\begin{equation}
  \begin{aligned}
    C_a(t)=&\frac{-x}{2\pi }\int_{R_{inner}}^{R_a}C(r, x)rdr,\label{eq:selfInteractionGeometry}
  \end{aligned}
\end{equation}
where
\begin{equation}
  C(r, x)=\frac{ \pi 4x^3}{\sqrt{2}\left(lm\right)^{3/2}}-\frac{2  \mathcal{E}\left(\frac{m-l}{m}\right)}{\sqrt{m} l},\label{eq:geometricFactor}
\end{equation}
with $R_a$ is the radius of the disk, $E$ is the (anti)neutrino energy, $l=(x-r)^2+x^2$, $m=(x+r)^2+x^2$, $\mathcal{E}(M)$ is the complete elliptic integral of the second kind, and $x$ is the distance from the center along the plane of the disk.

We take the fluxes of neutrinos and antineutrinos to have the spectrum of the Fermi Dirac flux with zero chemical potential so that 
for electron neutrinos or electron antineutrinos with temperature, $T$, the flux is 
\begin{equation}
\phi_{\nu_e(\bar{\nu}_e)}(E)= \frac{gc}{2\pi^2(\hbar c)^3}\frac{E^2}{1+e^{T/E}},
\end{equation}
where $g=1$ is the spin parameter.
For $\nu_\mu(\bar{\nu}_\mu)$ and $\nu_\tau(\bar{\nu}_\tau)$ with temperature, $T$, the flux is also taken to have a Fermi Dirac spectrum with zero chemical potential. However, 
for the case of the single radius model, the flux is rescaled by $f_0$,
\begin{equation}
  \phi_{a}(E)=f_0\frac{gc}{2\pi^2(\hbar c)^3}\frac{E^2}{1+e^{T/E}},
\end{equation}
where $a=\nu_\mu$, $\bar{\nu}_\mu$, $\nu_\tau$ or $\bar{\nu}_\tau$.
For the multiple radius model, the flux is not rescaled so $f_0=1$ and the amount of $\nu_\mu$, $\bar{\nu}_\mu$, $\nu_\tau$ and $\bar{\nu}_\tau$ flux is entirely determined by the disk radius and temperature.
Initially, the system has the same flux of $\nu\_\mu$, $\nu_\tau$, $\bar{\nu}_\mu$ and $\bar{\nu}_\tau$ from each flavor and these neutrinos all have the same emission geometry.
Thus the density matrices can be written as,
\begin{equation}
  \begin{aligned}
    \rho(t,E)=&\sqrt{2}G_F\left(
    \begin{array}{ccc}
      \phi_{\nu_e}(E)C_{\nu_e}(t) & 0 & 0 \\
      0&\phi_{\nu_\mu}(E)C_{\nu_\mu}(t) &0\\
      0&0& \phi_{\nu_\mu}(E)C_{\nu_\mu}(t) 
    \end{array}
    \right)\\
    \bar{\rho}(t,E)=&\sqrt{2}G_F\left(
    \begin{array}{ccc}
      \phi_{\bar{\nu}_e}(E)C_{\bar{\nu}_e}(t) & 0 & 0 \\
      0&\phi_{\nu_\mu}(E)C_{\nu_\mu}(t) &0\\
      0&0& \phi_{\nu_\mu}(E)C_{\nu_\mu}(t) 
    \end{array}
    \right).
  \end{aligned}
\end{equation}
At $t=0$, the part of the Hamiltonian that comes from the neutrino-self interaction, $H_{\nu\nu}$ (equation \ref{eq:defineHnunu}), has only one non-zero element which is the $ee^\text{th}$ element. We define this to be $V_{\nu}(t)$, 
\begin{equation}
  \begin{aligned}
    V_\nu(t)=& V_{\nu_e}(t) - V_{\bar{\nu}_e}(t)\\
     =& \sqrt{2}G_F\int_0^\infty \left(C_{\nu_e}(t)\phi_{\nu_e}(E)-C_{\bar{\nu}_e}(t)\phi_{\bar{\nu}_e}(E)\right)\,dE.\label{eq:defineVnu}
    \end{aligned}
\end{equation}
Sometimes it is instructive to examine the unoscillated potential, i.e. what the self-interaction potential would be if no oscillations were to occur.  We will use $V_\nu$ for this purpose.  Of course, as neutrinos oscillate, the self interaction potential does evolve. The $S$ matrices of Eq. \ref{eq:defineHnunu} gain off diagonal components and therefore so does $H_{\nu\nu}$.
The potential in the $ee^\text{th}$ component deviates from Eq. \ref{eq:defineVnu}, and we define 
\begin{equation}
V_{osc}=(H_{\nu\nu})_{ee} - Tr(H_{\nu\nu})/3. \label{eq:defineOscillatedVnu}
\end{equation}

Before calculating the neutrino flavor transformation above our two types of disk configurations, we compute the unoscillated self interaction potential, $V_\nu(t)$, using the values for $R_a$ and $T_a$ from Tables \ref{tab:singleDiskParameters} and \ref{tab:multiDiskParameters}.
We show the results in Figs \ref{fig:singleDiskPotentials} and \ref{fig:multiDiskPotentials} for the magnitudes of all the potentials, $\Delta_{12}$, $\left|\Delta_{32}\right|$, $V_e(t)$, and $\left|V_\nu(t)\right|$. 
Fig. \ref{fig:singleDiskPotentials} is computed for the single radius model (Table \ref{tab:singleDiskParameters}) and
Fig. \ref{fig:multiDiskPotentials} is computed for the multiple radius model (Table \ref{tab:multiDiskParameters}) .
The vacuum potentials, $\Delta_{12}$, and $\left|\Delta_{32}\right|$, are plotted with a dark green dotted line and light green finely dotted line respectively.  These vacuum potentials in the figure have
a neutrino energy of $E=15$ MeV. 
The self interaction potential, $\left| V_\nu(t)\right|$ is plotted in a dashed light blue line.  The horizontal axis in all plots is progress along the neutrino trajectory in cm.  Since we use trajectories that are at a 45 degree angle the progress along the trajectory is $\sim \sqrt{2} x$.

Close to the disk, at the start of the trajectory, the geometric effect on the self interaction potential varies little.
At these distances, where the scale of the disk is large compared to the distance of the neutrino from the disk, the geometric contributions of neutrinos and antineutrinos are similar: $C_{\bar{\nu}_e}\sim C_{\nu_e}$ and roughly constant.
The hotter antineutrino disk will contribute more flux ($\phi_{\bar{\nu}_e} > \phi_{{\nu}_e}$) and this
 higher antineutrino emission causes the self interaction potential to start out negative. 
Far from the disk, the geometric contribution to the self interaction potential has the form
\begin{equation}
  C_a(t) \sim \frac{3 (R_a^4-R_{inner}^4)}{256 \sqrt{2} x(t)^4}. \label{eq:farAwayGeometry}
\end{equation}
The turnover between these two behaviors is determined by the disk radius, $R_a$. 
In the single disk model of Fig. \ref{fig:singleDiskPotentials}, this turnover happens at about $x_{turn}\sim10^6$ cm because all flavors have the same geometry, with $R_a=4.5\times10^6$ cm for all $a$.

In the multiple disk model of Fig. \ref{fig:multiDiskPotentials}, a similar turnover in the potential would occur at about $x_{turn}\sim10^6$ cm as well.
However, because the disks for electron neutrinos and antineutrinos have different radii, there is an additional feature.
At $x_{sym}\sim10^6$ cm, the neutrino self interaction potential for the multiple radius disk shows a symmetric point where the self interaction potential goes through zero, at the place where the electron neutrino and electron antineutrino contributions to the self-interaction potential are the same.
This symmetric point is a consequence of the nontrivial geometric contribution to the self interaction potential (Eq. \ref{eq:selfInteractionGeometry}) and the differing electron neutrino and electron antineutrino radii and temperatures.

Again, the electron antineutrino disk is  
hotter than the electron neutrino disk, and the local emission surface produces more flux,
\begin{equation}
  \int_0^\infty \phi_{\nu_e}(E)\,dE < \int_0^\infty\phi_{\bar{\nu}_e}(E)\,dE ,
\end{equation}
regardless of the local behavior of the geometric factors, $C_{\bar{\nu}_e}(t)$ and $C_{\nu_e}(t)$.
Close to the disk surface, the disk sizes are less important in determining the self interaction potential (Eq. \ref{eq:defineVnu})  than the temperatures since the geometric contributions are essentially the same and constant. Therefore the self interaction potential will be negative and stay roughly constant close to the emission surface as can be seen in Figs. \ref{fig:singleDiskPotentials} and  \ref{fig:multiDiskPotentials}.

While the flux of the electron antineutrinos always dominates over the flux of the electron neutrinos everywhere on the trajectory, farther from the disk surface, as the disk sizes become important, the self interaction geometric contribution associated with the neutrinos becomes larger than the geometric contribution associated with the antineutrinos,  $C_{\bar{\nu}_e}(t) < C_{\nu_e}(t)$.  This occurs because the radii, $R_a$, that set the scale of the potentials' turnover are different: $R_{\bar{\nu}_e}<R_{\nu_e}$, so the  geometric contribution for the electron antineutrinos begins to decrease sooner than the geometric contribution for the electron neutrinos (Eq. \ref{eq:farAwayGeometry}).
The symmetric point at about $x_{sym}\sim10^6$ cm is the place where the two potentials are the same magnitude and sum to zero. Before this point $V_\nu$ is negative and after it becomes positive.

The matter potential, which is shown as the light blue line in Figs. \ref{fig:singleDiskPotentials} and \ref{fig:multiDiskPotentials}, depends on the mass density of material in the outflow and the 
electron fraction. The matter close to the disk where the outflow trajectory would be vertical 
stays at a relatively constant density, which we take to be close to $10^{10}$ g/cm$^3$. As the 
outflow enters the radial trajectory, where we begin our calculation, the matter expands, 
dropping the density  to roughly $10^{8}$ g/cm$^3$. After this turnover, as the mass streams far 
from the disk, it continues a decline in density, calculated assuming a constant mass outflow 
rate as described in Sec. \ref{sec:describeMergers}.

Before embarking on the neutrino flavor transformation 
calculation, we can use the (unoscillated) potentials plotted in Figs.
 \ref{fig:singleDiskPotentials} and  \ref{fig:multiDiskPotentials} to identify regions where various types of oscillation physics may take place.
In the single disk model, Fig. \ref{fig:singleDiskPotentials},
at the beginning of the trajectory, the matter potential, in a dark blue line, and the size of the self interaction potential, in the light blue line, are large.
The large self interaction potential corresponds to synchronized neutrino oscillations where all modes behave roughly the same way, even though they have different energies.
After the synchronized region, the matter potential and the neutrino self interaction potential have the same magnitude at $2\times 10^7$ cm creating a matter neutrino resonance region.
At about $10^8$ cm where the neutrino self interaction scale and the vacuum scale are roughly the same size, there is a nutation region, and at about $3\times 10^8$ cm, an MSW regime.
The multiple disk model, Fig. \ref{fig:multiDiskPotentials}, has these regions of interest at more or less the same places and an additional matter neutrino resonance region at $x_{sym}\sim10^6$ cm due to the symmetric point.

\section{Results}\label{sec:results}
We now wish to consider the oscillation of neutrinos for the models discussed in the previous 
two sections. We perform the calculations outlined in section \ref{sec:calculations} for each 
of our models with varying contributions from $\nu_\mu$, $\nu_\tau$, $\bar{\nu}_\mu$, and 
$\bar{\nu}_\tau$. We take the vacuum mixing parameters to be consistent with the current 
values of the Particle Data Group \cite{Agashe:2014kda}, 
$\theta_{12}=34.4^\circ,\theta_{13}=9^\circ, 
\theta_{23}=45^\circ,\Delta_{12}E=7.59\times10^{-5}$ eV$^2$, and 
$\left|\Delta_{23}E\right|=2.43\times10^{-3}$ eV$^2$. We use the inverted hierarchy although 
the matter-neutrino resonances are nearly hierarchy independent and we find very similar 
results for these transitions in the normal hierarchy. The self interaction Hamiltonian 
couples the neutrinos and antineutrinos of different energies together and our integration 
keeps track of 800 different neutrino energies between $1$ and $101$ MeV. The result of the 
calculation is the $S$ matrices, which we use to find the flux weighted survival 
probabilities,
\begin{equation}
  \begin{aligned}
    \left\langle P\right\rangle=&\frac{\int_0^\infty \phi_{\nu_e}(E)P_{\nu_e\to\nu_e}(E)\,dE}{\int_0^\infty\phi_{\nu_e}(E)\,dE}\\
    \left\langle\bar{P}\right\rangle=&\frac{\int_0^\infty \phi_{\bar{\nu}_e}(E)P_{\bar{\nu}_e\to\bar{\nu}_e}(E)\,dE}{\int_0^\infty \phi_{\bar{\nu}_e}(E)\,dE}.
  \end{aligned}
\end{equation}
The energy dependent survival probabilities and transition probabilities come from the $S$ matrix,
\begin{equation}
  \begin{aligned}
  P_{\nu_e\to\nu_e}(E)=&\left|S_{ee}(E)\right|^2\\
  P_{\nu_\mu\to\nu_e}(E)=&\left|S_{\mu e}(E)\right|^2\\
  P_{\nu_\tau\to\nu_e}(E)=&\left|S_{\tau e}(E)\right|^2\\
  \\
  P_{\bar{\nu}_e\to\bar{\nu}_e}(E)=&\left|\bar{S}_{ee}(E)\right|^2\\
  P_{\bar{\nu}_\mu\to\bar{\nu}_e}(E)=&\left|\bar{S}_{\mu e}(E)\right|^2\\
  P_{\bar{\nu}_\tau\to\bar{\nu}_e}(E)=&\left|\bar{S}_{\tau e}(E)\right|^2.
  \end{aligned}
\end{equation}
The evolution matrix, $\bar{S}$, for the antineutrinos obeys an equation similar to Eq. (\ref{eq:evolution}).

The capture rates for electron neutrinos and antineutrinos on neutrinos and protons respectively are approximated as
\begin{equation}
  \begin{aligned}
    \lambda_{\nu_e}=&\frac{\pi c G_F^2}{(\hbar c)^4}(c_V^2 + 3c_A^2 )\int_0^\infty \left(\phi_{\nu_e}^\prime(r,E)P_{\nu_e\to\nu_e}(E) + \phi_{\nu_\mu}^\prime(r,E)P_{\nu_\mu\to\nu_e}(E)+ \phi_{\nu_\tau}^\prime(r,E)P_{\nu_\tau\to\nu_e}(E)\right)(E + Q)^2\,dE\\
    \lambda_{\bar{\nu}_e}=&\frac{\pi c G_F^2}{(\hbar c)^4}(c_V^2 + 3c_A^2 )\int_{E_0}^\infty \left(\phi_{\bar{\nu}_e}^\prime(r,E)P_{\bar{\nu}_e\to\bar{\nu}_e}(E) + \phi_{\bar{\nu}_\mu}^\prime(r,E)P_{\bar{\nu}_\mu\to\bar{\nu}_e}(E) + \phi_{\bar{\nu}_\tau}^\prime(r,E)P_{\bar{\nu}_\tau\to\bar{\nu}_e}(E)\right)(E - Q)^2\,dE
  \end{aligned}
\label{eq:capturerates}
\end{equation}
where $Q=m_n-m_p$ is the nucleon mass difference and $E_0=(m_n-m_p+m_e)$ is the threshold energy for electron antineutrino capture.
The fluxes, $\phi^\prime_{a}(r,E)$, decrease appropriately as the distance from  the disk increases (this geometric effect is described in \cite{Surman:2003qt} and is different from $\phi_{a}(E)C_{a}$).  In addition to plotting survival probabilities we will plot the ratio of the capture rates in the case with neutrino oscillations to the case without neutrino oscillations.
In the absence of oscillation, the capture rates, $\lambda_{\nu_e}^0$ and $\lambda_{\bar{\nu}_e}^0$, are calculated with 
\begin{equation}
  \begin{aligned}
    P_{\nu_e\to\nu_e}(E)=&1\\
    P_{\nu_\mu\to\nu_e}(E)=&0\\
    P_{\nu_\tau\to\nu_e}(E)=&0\\
    \\
    P_{\bar{\nu}_e\to\bar{\nu}_e}(E)=&1\\
    P_{\bar{\nu}_\mu\to\bar{\nu}_e}(E)=&0\\
    P_{\bar{\nu}_\tau\to\bar{\nu}_e}(E)=&0,
  \end{aligned}
\end{equation} 
for all energies, $E$.

\subsection{Single Disk Models}\label{sec:singleDiskModels}

\begin{figure}
	\begin{subfigure}{0.45\textwidth}
		\centering
  		\includegraphics[angle=270,width=\textwidth]{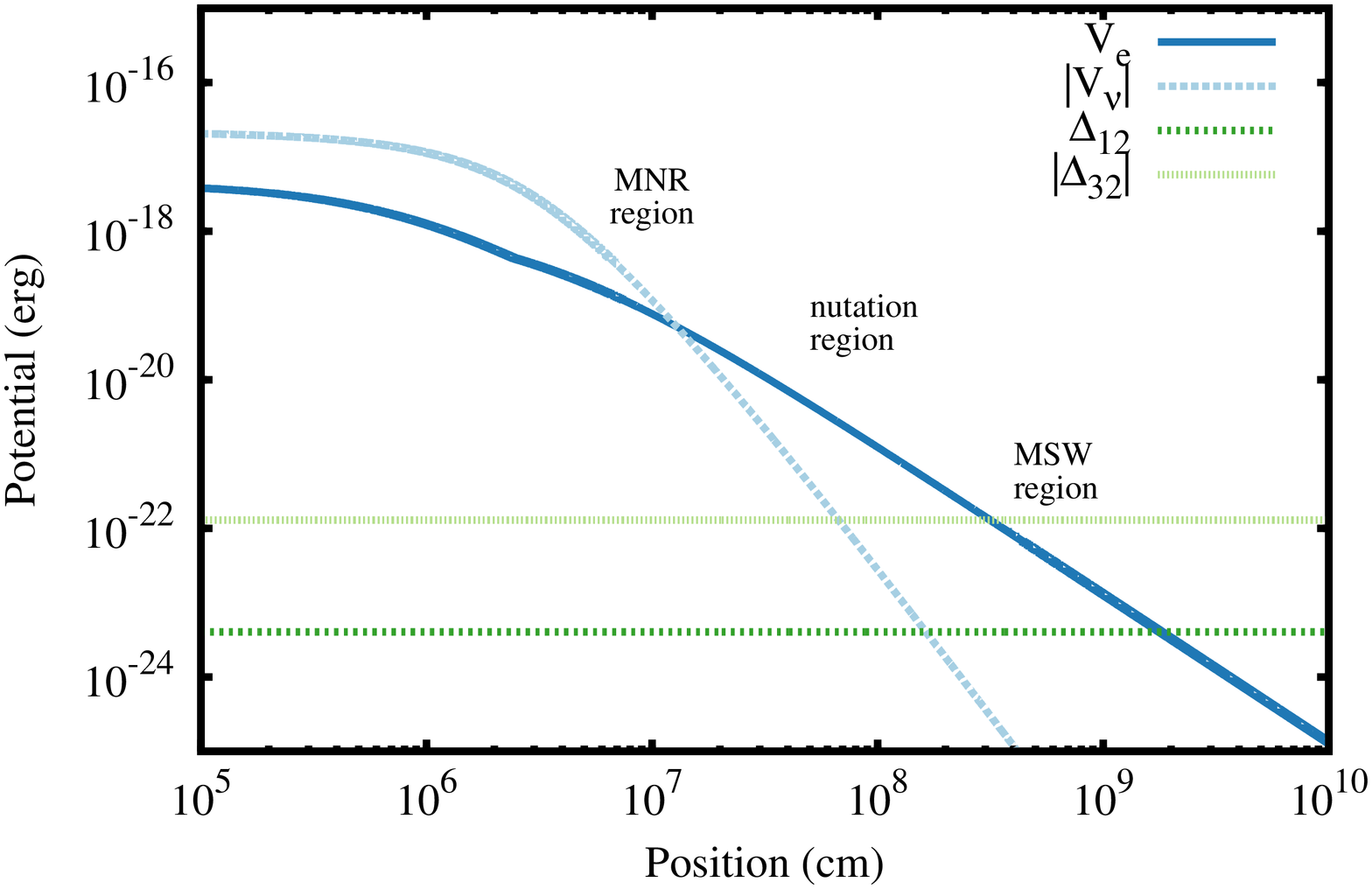}
\caption{}\label{fig:singleDiskPotentials} 
	  \end{subfigure}
  \hfill
  	 \begin{subfigure}{0.45\textwidth}
   		\centering
	  	\includegraphics[angle=270,width=\textwidth]{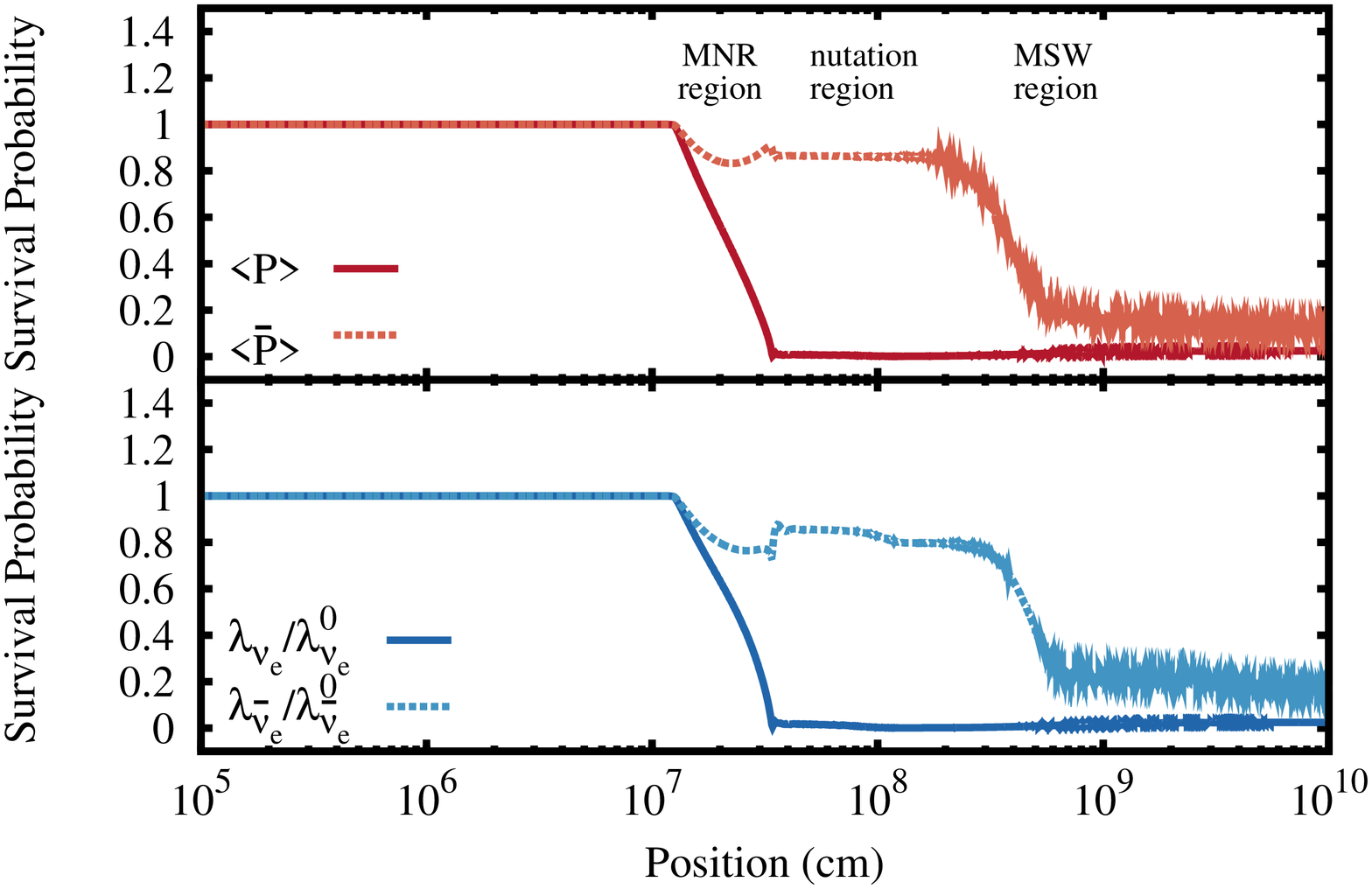}
\caption{}  
		\label{fig:nex024e2s20b2.0_0percent}   
	  \end{subfigure}
   \hfill
   	\begin{subfigure}{0.45\textwidth}
  		\centering
  		\includegraphics[angle=270,width=\textwidth]{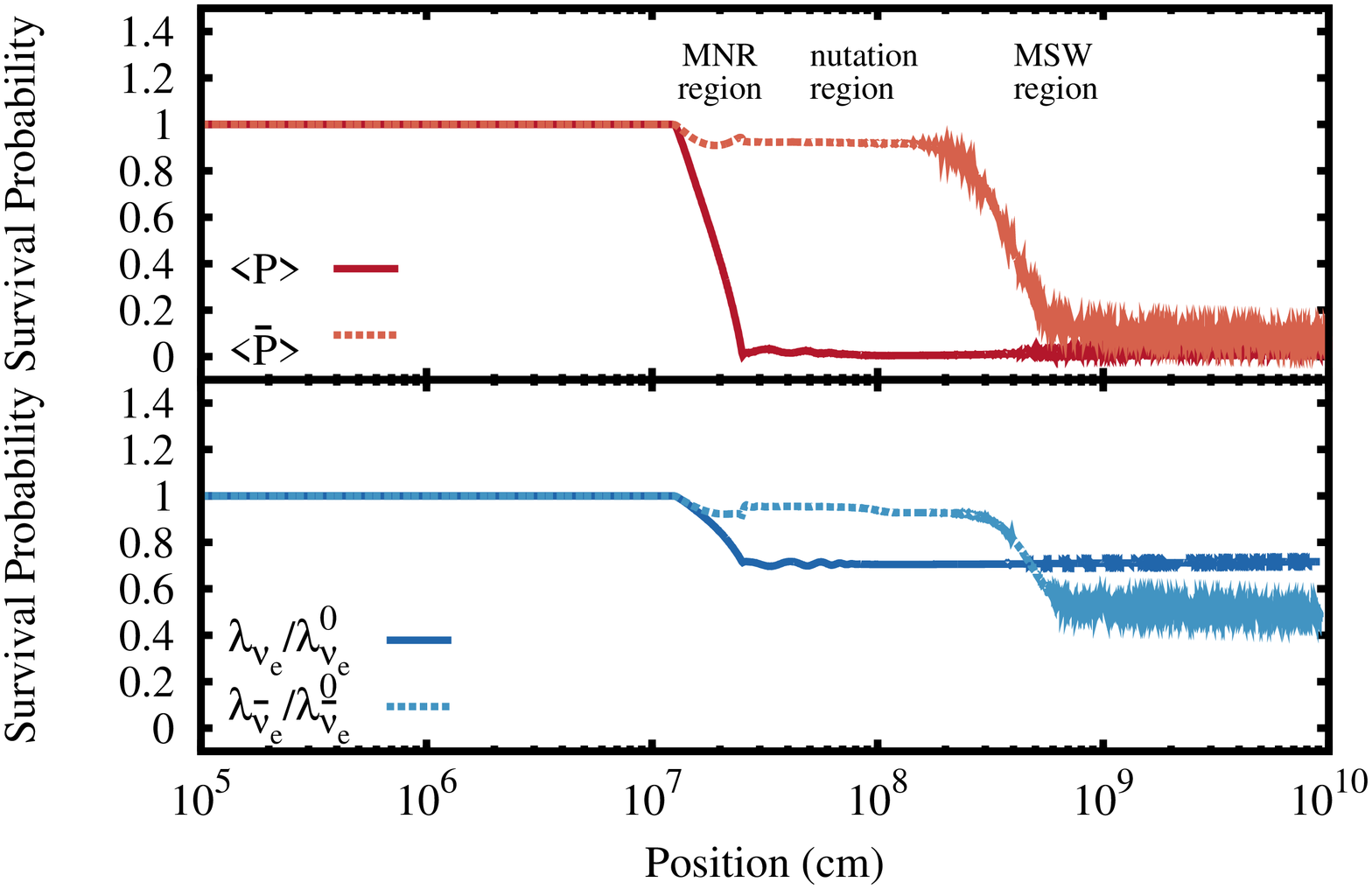}
\caption{}\label{fig:nex024e2s20b2.0_39percent}    
  	\end{subfigure}
   \hfill
   	\begin{subfigure}{0.45\textwidth}
  		\centering
  		\includegraphics[angle=270,width=\textwidth]{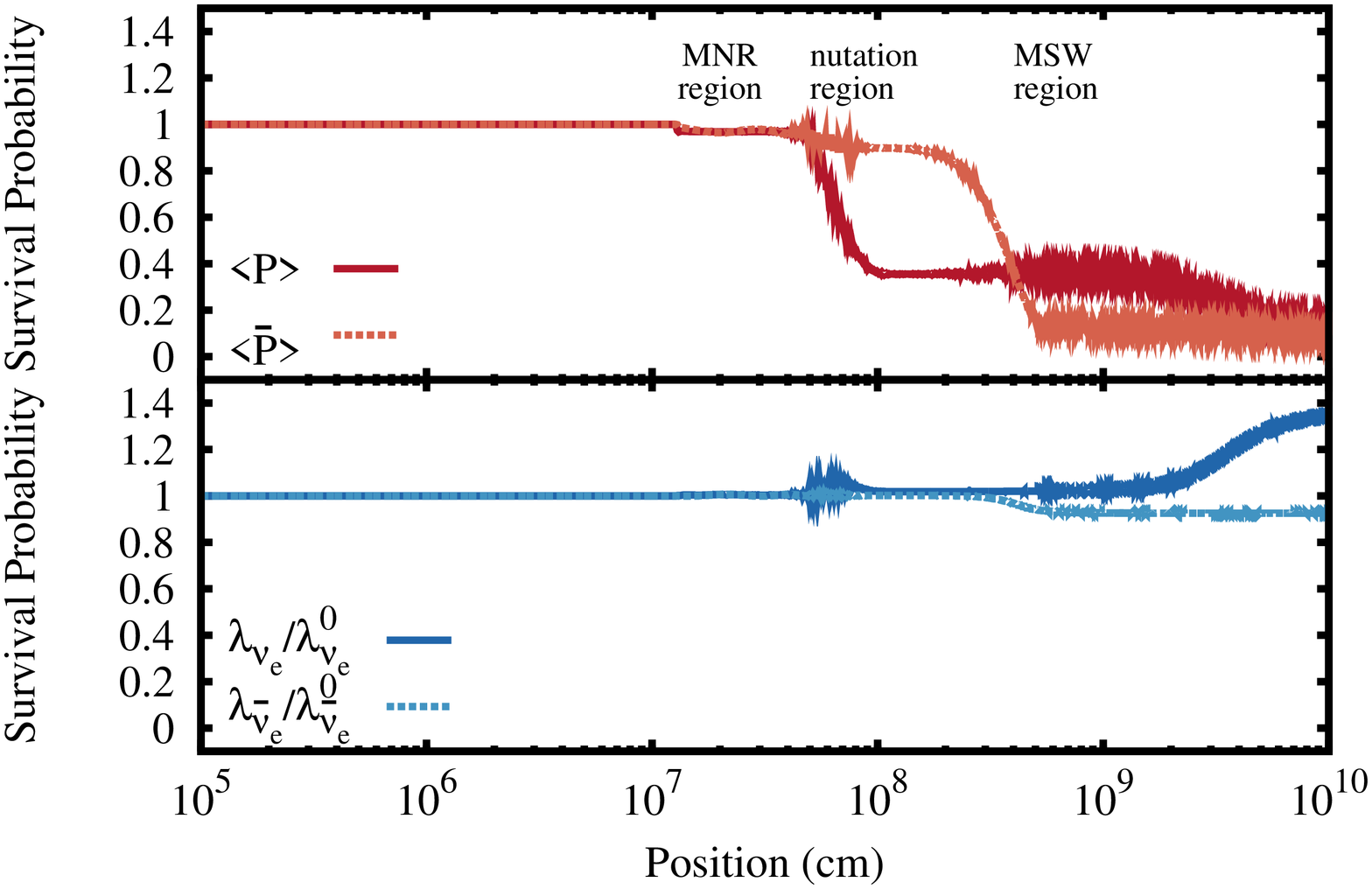}
\caption{}\label{fig:nex024e2s20b2.0_85percent}   
	\end{subfigure}
  \caption{\textbf{Single Disk Model:} All flavors of neutrinos and antineutrinos are emitted from a disk of the same size. 
  	The neutrino and antineutrino temperatures differ, as in Table \ref{tab:singleDiskParameters}. The horizontal axis in all plots is progress along the neutrino trajectory in cm. 
	\textit{Fig. \ref{fig:singleDiskPotentials}:} Potentials entering Hamiltonian from electrons as in Eq. \ref{eq:defineHe} and from neutrinos 
as in Eq. \ref{eq:defineVnu} in the absence of oscillation.
	 \textit{All other plots:} The top panel shows the flux weighted electron neutrino survival probability, $\left\langle P\right\rangle$ in red solid lines, 
	 and the flux weighted electron antineutrino survival probability, $\left\langle \bar{P}\right\rangle$, in dashed amber line.
	In the bottom panel, we show the relative capture rates of the electron neutrinos and antineutrinos.
	The relative capture rate of the neutrinos is a ratio of the electron neutrino capture rate when oscillations are taken into account, $\lambda_{\nu_e}$ to the electron neutrino capture rate when oscillations are not present, $\lambda_{\nu_e}^0$, and is shown as the dark blue line.
	The relative electron antineutrino capture rate is the analogous ratio, $\lambda_{\bar{\nu}_e}/\lambda_{\bar{\nu}_e}^0$, which we show in a light blue dashed line.
	\textit{Fig. \ref{fig:nex024e2s20b2.0_0percent}:} No mu or tau neutrinos are emitted from the disk.
	\textit{Fig. \ref{fig:nex024e2s20b2.0_39percent}:} Mu and tau neutrino fluxes are rescaled; $f_0$=0.35 relative to their blackbody fluxes.
	\textit{Fig. \ref{fig:nex024e2s20b2.0_85percent}:} Mu and tau neutrino fluxes are rescaled; $f_0$=0.75
	}
  \end{figure}

We first present the results of the calculations described in Sec. \ref{sec:calculations} using the single disk model discussed in Sec. \ref{sec:describeMergers}.
We calculate the neutrino oscillation pattern resulting from these potentials and show the results in Figs \ref{fig:nex024e2s20b2.0_0percent}, \ref{fig:nex024e2s20b2.0_39percent}, and \ref{fig:nex024e2s20b2.0_85percent} for different amounts of emitted mu and tau neutrinos and antineutrinos.

The potentials in Fig. \ref{fig:singleDiskPotentials} shape the oscillation regions that the neutrinos enter by their relative sizes.
The matter potential and the self interaction potential become the same size at $2\times 10^7$ cm.
We identify this position as a standard MNR region.
Standard MNR transitions 
have been explained in \cite{Malkus:2014iqa} and has similarities with the neutrino-antineutrino transformation discussed in  \cite{Vlasenko:2014bva}.
In a standard MNR transition,
the self interaction potential from Eq. (\ref{eq:defineOscillatedVnu}) changes so that it matches the size of the matter potential throughout the transition: $\left|V_e\right| \sim \left|V_{osc}\right|$.
This means that the on-diagonal component of $H$ stays near zero, i.e. 
\begin{equation}
V_{osc} +V_e+(H_V)_{ee}\sim0\label{eq:resonanceCondition}
\end{equation}
If the system is sufficiently adiabatic, a standard MNR transition will occur: neutrinos and antineutrino oscillate in just such a way that the potentials maintain the resonance.
The oscillated $V_{osc}$ changes to enforce Eq. \ref{eq:resonanceCondition}.
As discussed in \cite{Malkus:2014iqa}, the behavior of this resonance transition is well approximated by a single monoenergetic neutrino and antineutrino model.
In this context, \cite{Malkus:2014iqa} derived the diabaticity criteria that comes from matching the timescale of transition determined by the scale height, 
$\tau=\left|d(V_e/V_{\nu_e})/dl\right|^{-1}$, of the matter and self interaction potentials,
\begin{equation}
  \delta l_1=\tau\log\left(\frac{\alpha
+1}{\alpha
-1}\right),
\end{equation}
with the timescale determined by the capacity of system to change flavor, 
\begin{equation}
  \delta l_2=\frac{\alpha
}{\Delta \sin2\theta S_{ext}},
\end{equation}
where \
\begin{equation}
  \alpha
\equiv \frac{V_{\bar{\nu}_e}(t)}{V_{\nu_e}(t)}\label{eq:defineAlphaSymmetric}
\end{equation}
and 
$S_{ext}$ is the average difference between the $y$ components of Neutrino Flavor Isospin vectors for the neutrino and antineutrino \cite{Malkus:2014iqa}, which must be less than $(1+\alpha
)/2$.
Below this value, $\delta l_1=\delta l_2$ as long as $\theta$ is sufficiently large.
For the single disk model, the scale height is $\tau_{single}\sim 7\times10^6$ cm at the standard MNR region, where, $\alpha
\sim1.4$.
These values mean that in order for a transition to occur associated with $\delta m^2_{13}$, the mixing angle must be $\theta > 3 \times10^{-2}$. 
The measured mixing angle, $\theta_{13}\sim 0.2$, is well within that range.
The standard MNR regions in our models should be sufficiently adiabatic to sustain transitions. 

We now look at our numerical calculations to see if the transitions indeed occur.  The results of these calculations are shown in Figs. \ref{fig:nex024e2s20b2.0_0percent}, \ref{fig:nex024e2s20b2.0_39percent} and \ref{fig:nex024e2s20b2.0_85percent}.  If there are no mu or tau neutrinos and antineutrinos, the transition results in flux weighted survival probability for electron neutrinos dropping to nearly zero and the flux weighted survival probability of electron antineutrinos returning to nearly one, as in Fig. \ref{fig:nex024e2s20b2.0_0percent}.  This is the characteristic behavior of a standard MNR described in \cite{Malkus:2014iqa}.  The relative capture rates of electron neutrinos are shown in solid dark blue and electron antineutrinos are shown in light dashed blue lines in the lower panel. 
Because no mu or tau neutrinos are present initially, these relative capture rates track closely to the weighted survival probabilities. 

Increasing the initial amount of mu and tau neutrinos by a small amount has little effect on the survival probabilities, but a significant effect on the capture rates. This can be seen in Fig. \ref{fig:nex024e2s20b2.0_39percent} where a small quantity of mu and tau neutrinos are now emitted from the disk.  The number flux of each of these types of neutrinos and neutrinos is slightly under $40\%$ 
of the electron antineutrino flux.
We see from a comparison of the top panel of Fig. \ref{fig:nex024e2s20b2.0_0percent} and Fig. \ref{fig:nex024e2s20b2.0_39percent} that the addition of the mu and tau neutrinos at this modest level changes the survival probabilities very little.  However, a comparison of the bottom panels shows that the relative capture rates have changed.
Both the relative capture rates of electron neutrinos and of the electron antineutrinos begin at 1, like the weighted survival probability.  During the MNR, the capture rate of electron neutrinos drops only to about 70\%. The small change in capture rate occurs despite the fact that the MNR results in a strong transition from initially electron flavor neutrinos to other flavors,  because mu and tau neutrinos {\em also} transform at the MNR to electron flavor.

Increasing the initial amount of mu and tau neutrinos by a larger amount than in  Fig. \ref{fig:nex024e2s20b2.0_39percent}, has a more dramatic effect. In Fig. \ref{fig:nex024e2s20b2.0_85percent}, where the mu and tau neutrino flux is $85\%$ of the electron antineutrino component, the top panel shows that neutrinos initially in electron flavor remain in electron flavor, even as they pass through the MNR region. 
The bottom panel of Fig. \ref{fig:nex024e2s20b2.0_85percent} is consistent with this failure of the MNR to result in transition; the capture rate of electron neutrinos remains the same, just as if there were no MNR region.  This behavior occurs generically when the mu and tau fluxes approach the same level as the electron neutrinos and antineutrinos.  During MNR transitions, the neutrinos and antineutrinos transform in such a way so that the self interaction potential matches the matter potential, Eq. \ref{eq:resonanceCondition}.  But since a neutrino flavor transformation involves a trading of electron type neutrinos with mu and tau type neutrinos, if there are similar numbers of electron neutrinos as other types then the self-interaction potential, Eq. \ref{eq:defineOscillatedVnu}, cannot change much and  no transformation occurs. 

After the MNR region, the system passes through a nutation region, which happens at about $10^8$ cm in Figs. \ref{fig:nex024e2s20b2.0_0percent}, \ref{fig:nex024e2s20b2.0_39percent}, and \ref{fig:nex024e2s20b2.0_85percent}.
Nutation oscillations are a high neutrino density phenomenon that have been deeply studied in the supernova setting, e.g. \cite{Duan:2005cp}.  They have also been studied in the disk setting, \cite{Malkus:2012ts, Dasgupta:2008cu}. They occur when the self interaction potential approaches the size of the vacuum scale and a required initial condition is that the electron neutrinos and antineutrinos begin approximately in their flavor eigenstates.  During the transition the neutrinos and antineutrinos are said to be ``locked'' in that their transition probabilities are linked.  The transitions are largest in the inverted hierarchy which is the case for the calculations presented here.  In Figs. \ref{fig:nex024e2s20b2.0_0percent}, \ref{fig:nex024e2s20b2.0_39percent}, and \ref{fig:nex024e2s20b2.0_85percent} the nutation region occurs after the matter neutrino resonance region which would be a typical expectation in a disk system.  These figures show that nutation oscillations occur only in the case where MNR transitions do not, i.e Fig. \ref{fig:nex024e2s20b2.0_85percent}.  This is because the MNR transition moves the electron neutrinos and antineutrinos out of their initial eigenstates, and the required initial conditions for a nutation oscillation are not met.  We note that nutation oscillations have been shown to depend upon multiangle effects which are expected to be large in a disk setting.  The authors of 
\cite{Dasgupta:2008cu} discussed a formalism by which to use a single angle calculation to find nutation oscillation survival probabilities that are robust in a merger.

A Mikeyev Smirnov and Wolfenstein (MSW) region occurs when the matter potential becomes the same size as the vacuum scale. In the models of Figs \ref{fig:nex024e2s20b2.0_0percent}, \ref{fig:nex024e2s20b2.0_39percent}, and \ref{fig:nex024e2s20b2.0_85percent}, this region is at about $4\times 10^8$ cm. 
At the densities where MSW regions occur, the self interaction potential tends to be small and the neutrinos and antineutrinos of different energies act almost independently.
The energy dependent oscillations will then occur only if the system is sufficiently adiabatic.
In all three examples, we see that the MSW regime results in transitions of the electron antineutrinos to other flavors. These transitions of electron antineutrinos 
would not be possible in the normal hierarchy where instead electron neutrinos would transform.

\subsection{Multiple Disk Model}\label{sec:multidisk}

\begin{figure}
	\begin{subfigure}{0.45\textwidth}
		\centering
		  \includegraphics[angle=270,width=\textwidth]{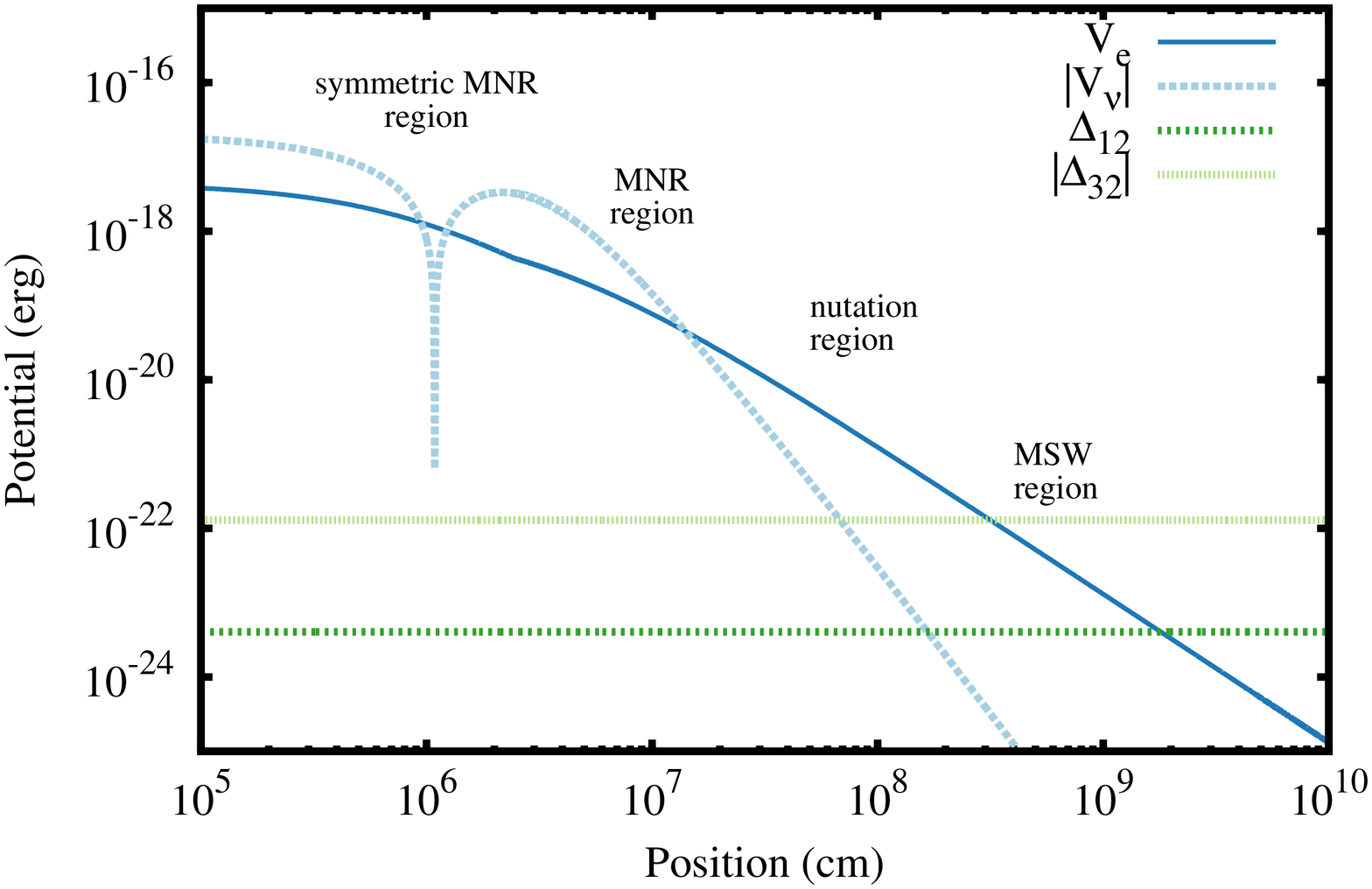}
\caption{}\label{fig:multiDiskPotentials} 
  	\end{subfigure}
   \hfill
   	\begin{subfigure}{0.45\textwidth}
  		\centering
  \includegraphics[angle=270,width=\textwidth]{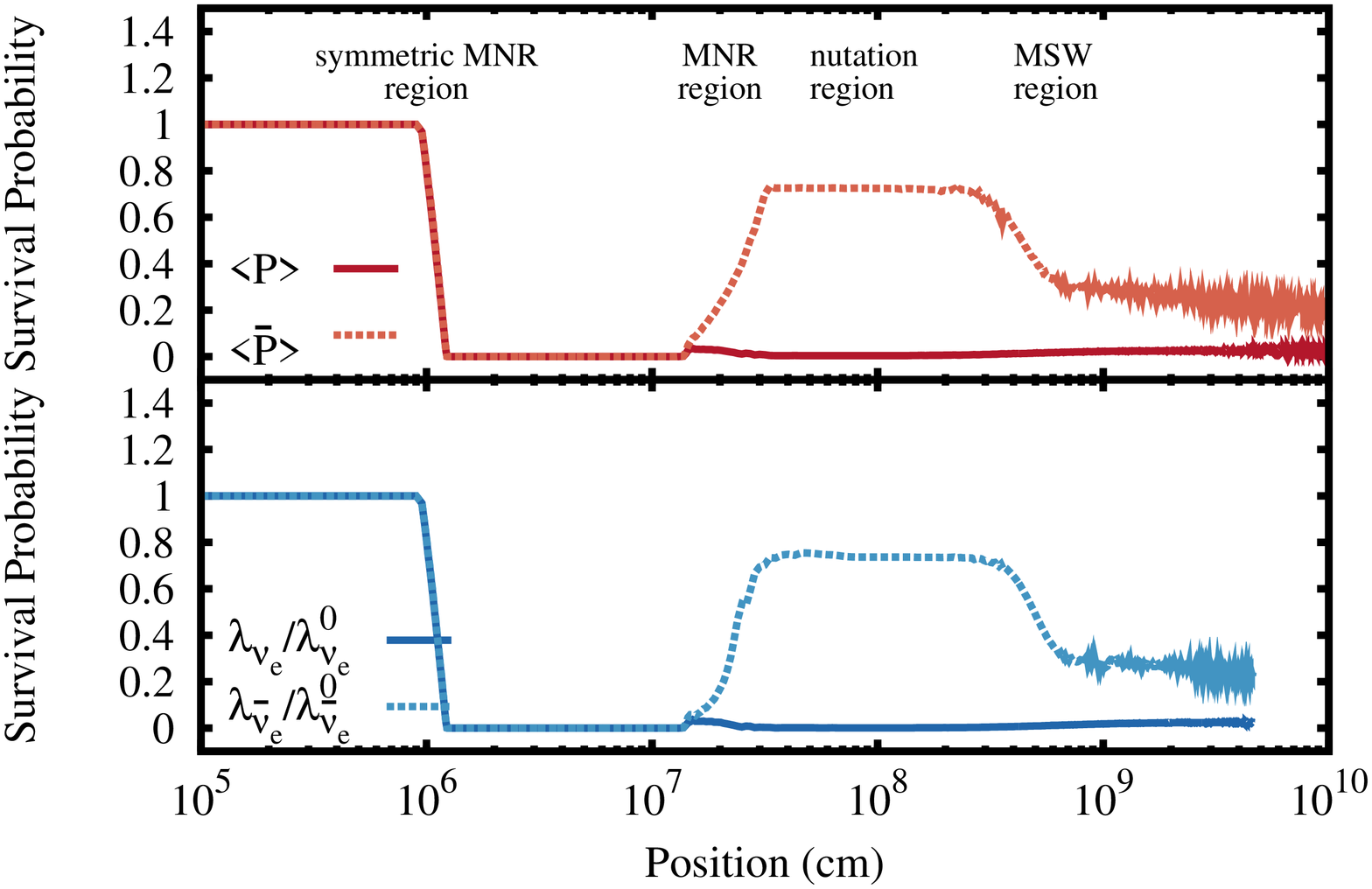}
\caption{}\label{fig:nex024e2s20b2.0MultiNoHeavies}
  	\end{subfigure}
   \hfill
   	\begin{subfigure}{0.45\textwidth}
  		\centering
  		\includegraphics[angle=270,width=\textwidth]{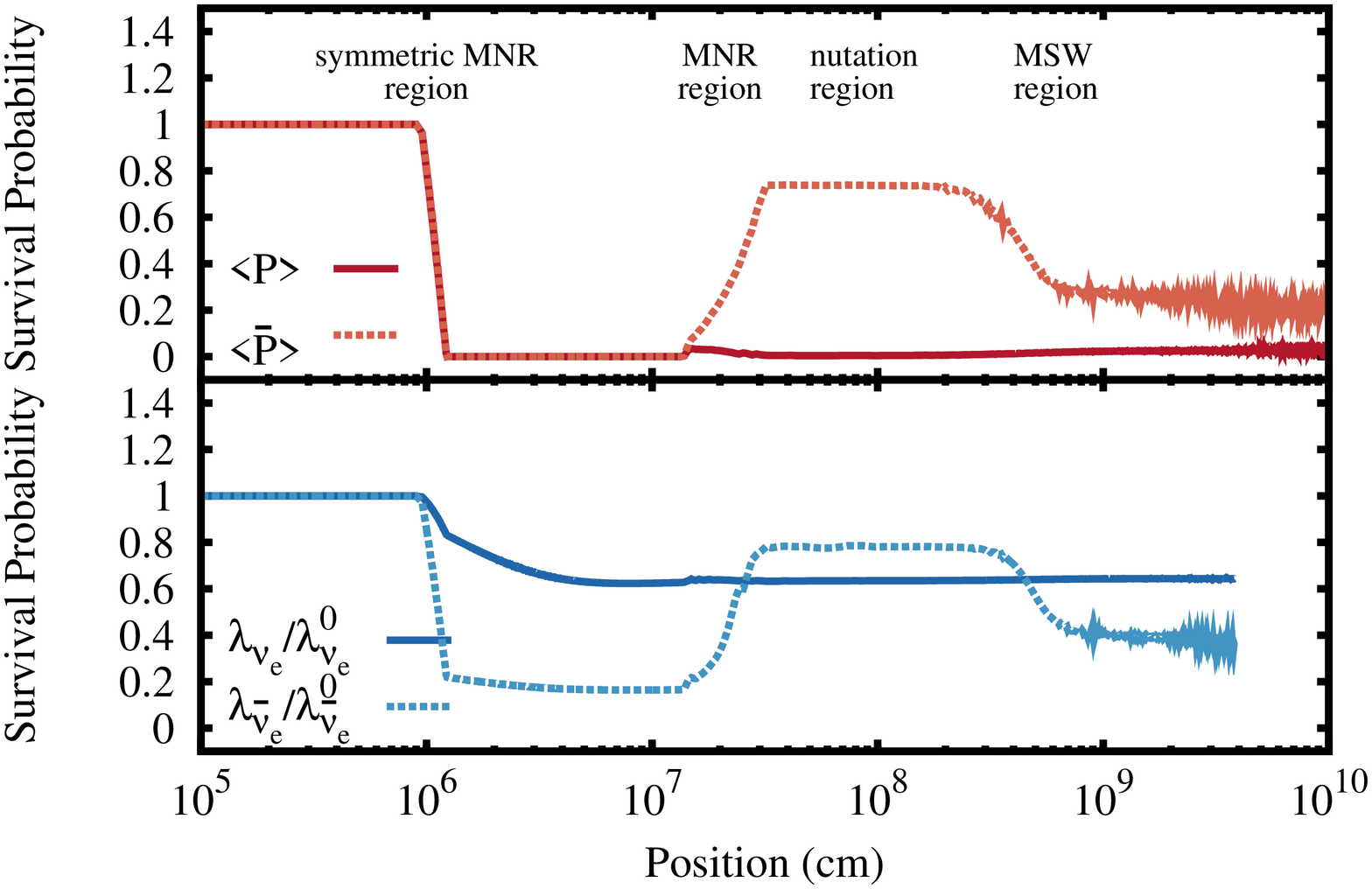}
\caption{}\label{fig:nex024e2s20b2.0Multi20Percent}
  	\end{subfigure}
   \hfill
   	\begin{subfigure}{0.45\textwidth}
  		\centering
		  \includegraphics[angle=270,width=\textwidth]{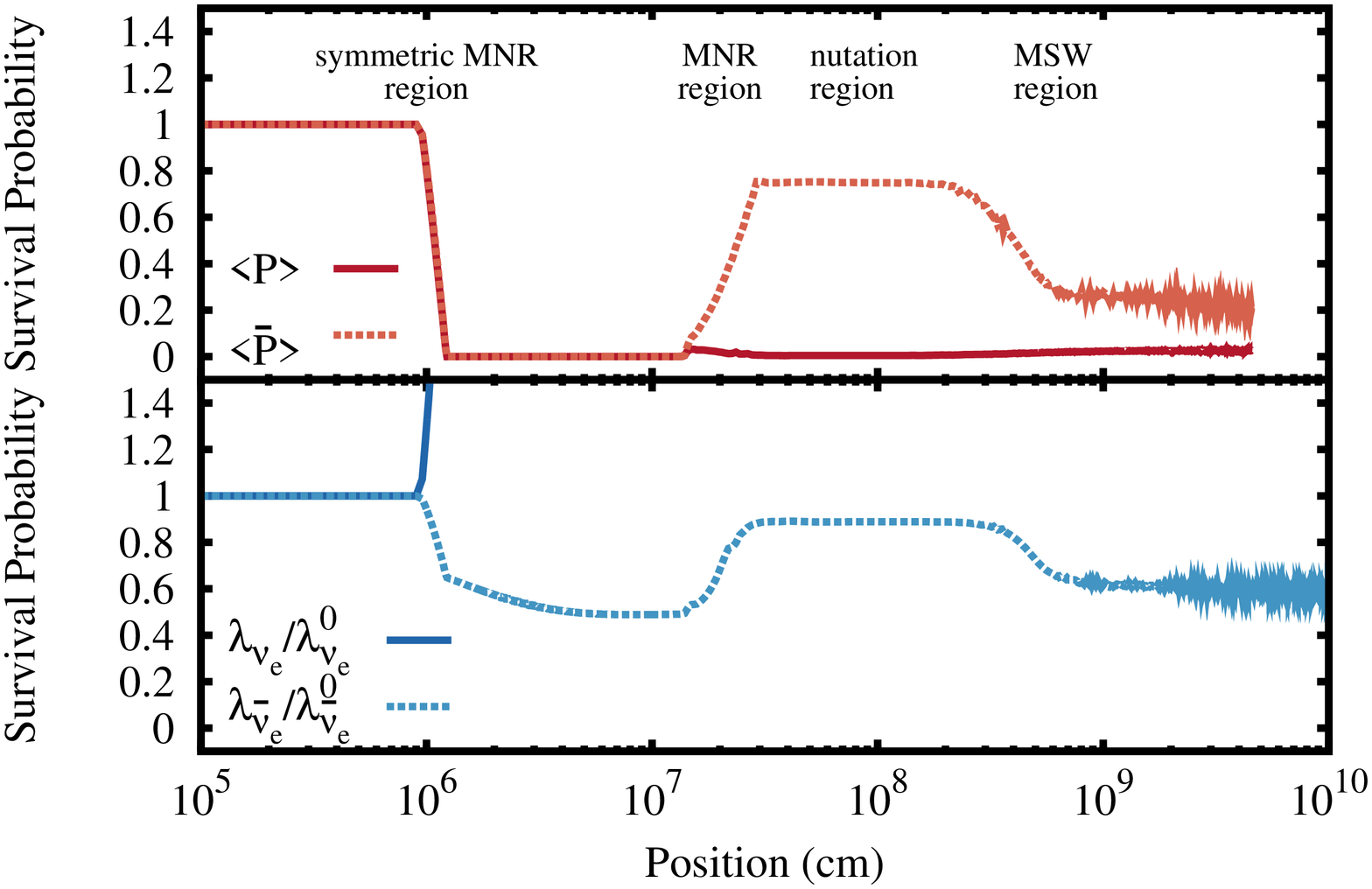}
\caption{}\label{fig:nex024e2s20b2.0Multi65Percent}
	\end{subfigure}
\caption{\textbf{Multiple Disk Model} Different flavors of neutrinos and antineutrinos are emitted from disks of different sizes, as in Table \ref{tab:multiDiskParameters}, i.e. $R_{\nu_e}= 5.2\times10^6$ cm and $R_{\bar{\nu}_e}=3.9\times10^6$ cm. The horizontal axis in all plots is progress along the neutrino trajectory in cm. 
	\textit{Fig. \ref{fig:multiDiskPotentials}:} Potentials entering Hamiltonian from electrons as in Eq. \ref{eq:defineHe} and from neutrinos 
as in Eq. \ref{eq:defineVnu} in the absence of oscillation.
	 \textit{All other plots:} The top panel shows the flux weighted electron neutrino survival probability, $\left\langle P\right\rangle$ in red solid lines, 
	 and the flux weighted electron antineutrino survival probability, $\left\langle \bar{P}\right\rangle$, in dashed amber line.
	In the bottom panel, we show the relative capture rates of the electron neutrinos and antineutrinos.
	The relative capture rate of the neutrinos is a ratio of the electron neutrino capture rate when oscillations are taken into account, $\lambda_{\nu_e}$ to the electron neutrino capture rate when oscillations are not present, $\lambda_{\nu_e}^0$, in a dark blue line.
	The relative electron antineutrino capture rate is the analogous ratio, $\lambda_{\bar{\nu}_e}/\lambda_{\bar{\nu}_e}^0$, which we show in a light blue dashed line.
	\textit{Fig. \ref{fig:nex024e2s20b2.0MultiNoHeavies}:} No mu and tau neutrinos or antineutrinos are emitted.
	\textit{Fig. \ref{fig:nex024e2s20b2.0Multi20Percent}: } Mu and tau neutrinos and antineutrinos are emitted from a disk with radius of $1.8\times 10^6$ cm.
	\textit{Fig.\ref{fig:nex024e2s20b2.0Multi65Percent}:} Mu and tau neutrinos and antineutrinos are emitted from a disk with radius of $2.4\times 10^6$ cm.}
 \end{figure}

In the multiple disk model, neutrinos of different flavors are emitted from disks of different radii as given in Table \ref{tab:multiDiskParameters}.  There are three different emission surfaces, one for electron neutrinos, one for electron antineutrinos and one for all other flavors.  The potentials for this model are shown in Fig. \ref{fig:multiDiskPotentials} and the different regions of oscillation behavior were discussed in Sec. \ref{sec:calculations}.   In this section we discuss the results of the multiple disk model calculations, which are shown in Figs. \ref{fig:nex024e2s20b2.0MultiNoHeavies}, \ref{fig:nex024e2s20b2.0Multi20Percent}, and \ref{fig:nex024e2s20b2.0Multi65Percent}.

Initially, the multiple disk models behave in the same was as the single disk models and have 
a region of synchronized behavior close to the disk, so that the survival probability is essentially one in the region before
$\sim 10^6$ cm.  However, after this point the multiple disk models start to exhibit a differences with the single disk models because they enter the symmetric MNR region which does not exist in the single disk model.  In the symmetric MNR region 
 $V_{osc}$ from Eq. \ref{eq:defineOscillatedVnu} becomes the same size as $V_e$.
As discussed in section \ref{sec:calculations}, in the multiple disk models, the difference in the changing geometric
factors causes the self interaction potential $V_\nu$, that would otherwise have large magnitude,  to sweep through zero.  We call this a symmetric resonance region because it occurs close to the point where $\alpha = 1$.  The  transitions in this region are similar to those observed in \cite{Malkus:2012ts} and the phenomenology of these transitions has similarities to the standard MNR \cite{Malkus:2014iqa}. In all three of Figs.  \ref{fig:nex024e2s20b2.0MultiNoHeavies}, \ref{fig:nex024e2s20b2.0Multi20Percent}, and \ref{fig:nex024e2s20b2.0Multi65Percent}, both neutrinos and antineutrinos undergo a nearly complete transformation.

The mu and tau  neutrinos  can prevent the MNR transition if their flux becomes roughly the same as the electron neutrinos and antineutrinos.  This case was shown in the single disk model in Fig. \ref{fig:nex024e2s20b2.0_85percent}.  However, in all the examples considered as multiple disk models, the mu and tau neutrinos are either non-existent as in Fig.  \ref{fig:nex024e2s20b2.0MultiNoHeavies} or emitted from a smaller disk than the $\nu_e$ and $\bar{\nu}_e$, $R_{\nu_{\mu}} = 1.8\times 10^6$ cm in Fig. \ref{fig:nex024e2s20b2.0Multi20Percent}  and $R_{\nu_\mu} =2.4 \times 10^6$ cm in Fig. \ref{fig:nex024e2s20b2.0Multi65Percent}.  Therefore they do not give roughly the same contribution as the electron type neutrinos and antineutrinos at the position of the resonance and the transition can proceed.  

While the presence of a relatively modest number of  mu and tau neutrinos does not effect the survival probability of the electron neutrinos and antineutrinos at the symmetric matter neutrino resonance, they do alter the capture rates.  This is again because the electron neutrinos don't simply ``oscillate away'', instead they exchange places with the mu/tau type.  For example,
the lower panel of Fig. \ref{fig:nex024e2s20b2.0Multi20Percent} shows 
in the blue solid line  the relative capture rate of electron neutrinos, which drops from one to only about 0.6 at the symmetric matter neutrino resonance transition.
In the lower panel of Fig. \ref{fig:nex024e2s20b2.0Multi65Percent}, where the mu and tau contributions are a little larger, the electron neutrino capture rate is {\it higher} than had there been no transformation at all.  This is due to the higher energy of the mu and tau neutrinos at emission and the energy squared dependence of the neutrino capture cross section.

After the symmetric matter neutrino resonance transition, the neutrinos encounter 
a standard MNR region at about $2\times10^7$ cm in Figs \ref{fig:nex024e2s20b2.0MultiNoHeavies}, \ref{fig:nex024e2s20b2.0Multi20Percent}, and \ref{fig:nex024e2s20b2.0Multi65Percent}. 
Whether or not a standard MNR transition occurs depends on the state of the system as it enters the standard MNR region.
Transitions cannot occur unless the resonance condition, Eq. \ref{eq:resonanceCondition}, is fulfilled and for this to happen, the neutrino self interaction potential must be negative.
While the potential, $V_\nu$ begins negative close to the disk, the changing geometric factors cause a relative sign change in the potential (see Eq. \ref{eq:defineVnu}) so that $V_\nu$ is positive at the position of the standard MNR.  However, $V_\nu$ is what the self-interaction potential would be in the absence of oscillations. The system has already undergone a transformation at the symmetric MNR and this transition produces
an {\em additional} change in the sign so that the actual self-interaction potential, $V_{osc}$, is negative.
Thus if a symmetric MNR transition occurs, the system is set up
up favorably for a standard MNR transition.  Since in all our examples a symmetric MNR transition occurs, we expect a standard MNR  transition as well. The change in survival probability at the standard MNR region can be seen in the top panels of  Figs. \ref{fig:nex024e2s20b2.0MultiNoHeavies}, \ref{fig:nex024e2s20b2.0Multi20Percent}, and \ref{fig:nex024e2s20b2.0Multi65Percent} to follow the typical behavior of a standard MNR transition.  Similar to the single disk model, the standard MNR transitions have a non trivial effect on the capture rates as can be seen in the bottom panels of these figures.

After the MNR regions, the neutrinos encounter the nutation region at about $10^8$ cm.  As in the single disk models, Figs. \ref{fig:nex024e2s20b2.0_0percent} and \ref{fig:nex024e2s20b2.0_39percent},  the system is no longer in the original flavor eigenstates and the nutation region results in no flavor transformation. The neutrinos then enter an MSW region after several times $10^8$ cm, where some flavor transformation takes place in Figs \ref{fig:nex024e2s20b2.0Multi20Percent}, \ref{fig:nex024e2s20b2.0Multi20Percent}, and \ref{fig:nex024e2s20b2.0Multi65Percent}.
Also, as in the single disk case, since we are using the inverted hierarchy, the electron antineutrinos in the multiple disk case transform to other flavors at the MSW region.

\section{Nucleosynthesis}\label{sec:nucleosynthesis}

\begin{figure}
		  \includegraphics[width=0.5\textwidth]{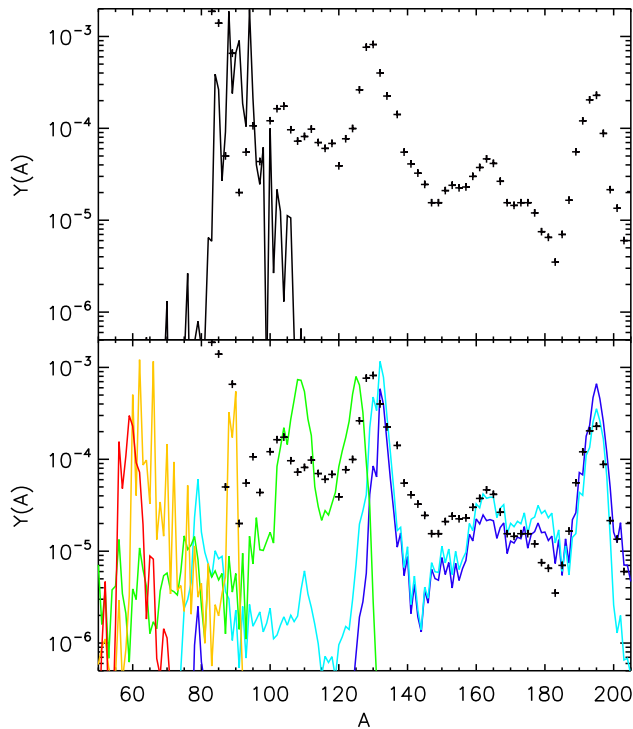}
		  \caption{Nucleosynthesis resulting from the multiple disk models. 
		  The vertical axis shows abundance and the horizontal axis shows the mass number A of elements produced.
		 \textit{Both Panels}: The black pluses show  scaled solar  $r$-process residuals. 
		\textit{Top Panel:} Production from the multiple disk model with no oscillations is shown in black.
		\textit{Bottom Panel:} Production from the multiple disk model with oscillation that includes no initial mu and tau neutrinos is shown in dark blue.
	       Oscillation calculations where mu and tau neutrinos have a flux relative the electron antineutrinos  at 5\%  (light blue), at 10\% (green), at 20\% (yellow) and at 65\% (red) are shown.
	       }\label{fig:rprocessAbundance} 
\end{figure}

The symmetric MNR region shown in Fig. \ref{fig:multiDiskPotentials} occurs quite close to the 
disk and thus will impact the element synthesis in outflowing material from the inner disk 
regions. This material starts out as primarily free nucleons, where neutrons far outnumber 
protons, and as it moves outward the composition evolves via the weak interactions
\begin{eqnarray}
\nu_{e}+n & \rightleftharpoons & p+e^{-} \\
\bar{\nu}_{e}+p & \rightleftharpoons & n+e^{+}.
\label{eqn:weak}
\end{eqnarray}
The approximate rates for the forward reactions are given in Eq. \ref{eq:capturerates}. In the nucleosynthesis calculations described in this section we include also the weak magnetism contribution \cite{Horowitz:2001xf} to these rates.
Since the disk emits more antineutrinos than neutrinos and the antineutrinos tend to be hotter, 
neutron-rich outflows are favored; the outflows may also retain some of the neutron-richness of 
the disk. As the material expands and cools, the protons are quickly bound into alphas, and of 
the weak reactions above only the top forward reaction, $\nu_{e}+n \rightharpoonup p+e^{-}$, 
continues to operate. The protons thus produced are promptly bound into alphas as well. At this 
stage of the nucleosynthesis neutrinos act to increase the number of seed nuclei and reduce the 
number of free neutrons available for capture on the seeds. This is called the `alpha effect' 
\cite{Fuller+95,Meyer+98}, and it limits how far in $A$ the nucleosynthesis can proceed. Most 
calculations of merger disk outflow nucleosynthesis 
\cite{Surman:2008qf,Metzger+08,Wanajo:2011vy,Caballero+12,Wanajo:2014wha,Just:2014fka,Perego:2014fma} favor 
production of weak ($80<A<120$), rather than main ($A>120$), $r$-process nuclei. The MNR can 
potentially alter this conclusion.

The influence of an MNR transition close to the emission surface 
 on outflow nucleosynthesis was first pointed out for a 
collapsar-type disk in \cite{Malkus:2012ts}. A collapsar disk emits primarily electron flavor 
neutrinos and antineutrinos, so when the MNR transition occurs it is (from the perspective of the nuclear matter in the 
vicinity) as if the neutrinos disappear. If this happens in the region of the outflow where 
alphas are forming, the alpha effect can be cut off completely. With fewer seeds formed and 
more free neutrons remaining, a vigorous main $r$ process can result \cite{Malkus:2012ts}.

\begin{figure}
	\includegraphics[width=0.5\textwidth]{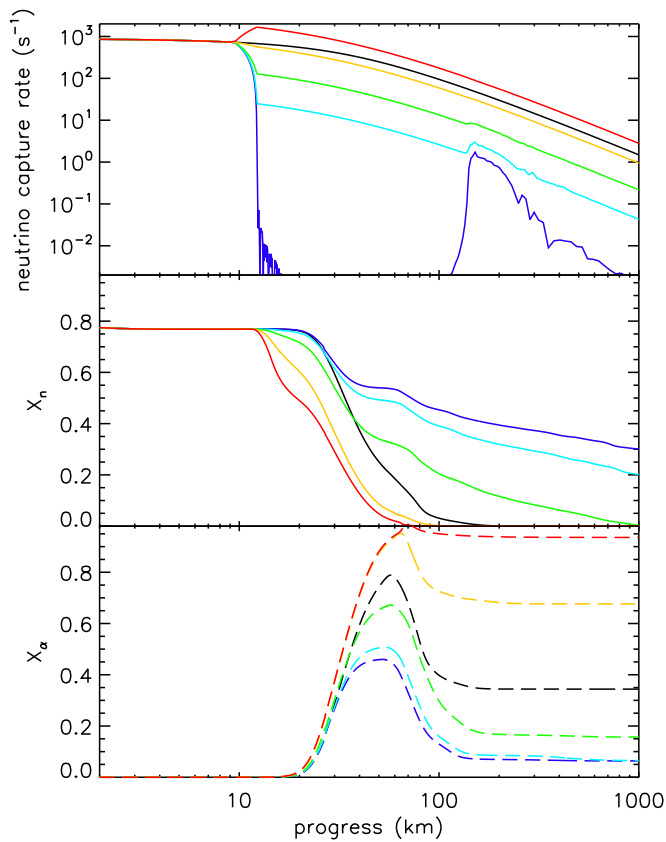}
	\caption{
	\textit{Top Panel:} Neutrino capture rates from the multiple disk models as a function of position along the outflow trajectory. 	
	\textit{Middle Panel:} Neutron mass fraction as a function of position along the outflow trajectory.
	\textit{Bottom Panel:} Alpha particle mass fraction as a function of position along the trajectory.
		\textit{All Panels} The multi disk model with no oscillations is shown in black.
		The multiple disk model with oscillation that includes no initial mu and tau neutrinos is shown in dark blue.
	       Oscillation calculations where mu and tau neutrinos are included at 5\% are shown in light blue, at 10\% in green, at 20\% in yellow and at 65\% in red.
	       }\label{fig:rprocessProgress} 
\end{figure}

Exactly how the MNR will influence merger outflow nucleosynthesis depends on the relative 
amounts of electron and mu/tau flavors emitted. If there is little mu/tau emission, we expect an effect similar to that described above for the collapsar case. This is shown in the 
 dark/light blue lines in Figs. \ref{fig:rprocessAbundance} and \ref{fig:rprocessProgress}. Here, we start with the neutrino emission 
from the multiple disk example, described in Sec. \ref{sec:multidisk}, and calculate the element synthesis as 
in \cite{Surman:2008qf,Malkus:2012ts} along the outflow trajectory described in Sec. \ref{sec:describeMergers}. The top 
panel of Fig. \ref{fig:rprocessAbundance} shows the final abundances for the case with no neutrino oscillations, 
compared to the solar $r$-process abundance pattern. Primarily $A\sim 80-90$ neutron-rich 
nuclei are produced. The material starts out very neutron-rich, as shown in the middle panel of 
Fig. \ref{fig:rprocessProgress}, but the alpha effect limits the resulting nucleosynthesis to a weak $r$ process. When 
the neutrino oscillations illustrated in Fig. 3 are included in the calculation, we find the 
MNR can radically change this picture. Fig.
\ref{fig:nex024e2s20b2.0MultiNoHeavies} 
shows that the symmetric MNR region occurs at a point about 10 km along the outflow trajectory. If there are little to no 
mu/tau neutrinos to oscillate into the electron flavors, the electron neutrino capture rate 
will drop steeply in the MNR region. This is shown for example cases with little (light blue 
lines) to no (dark blue lines) mu/tau emission in the top panel of Fig. \ref{fig:rprocessProgress}. For the outflow 
conditions considered here, MNR occurs just before the alphas start forming, as depicted in the bottom panel of Fig. \ref{fig:rprocessProgress}. Thus, the alpha effect is completely removed by the MNR ---fewer 
alphas form (bottom panel of Fig. \ref{fig:rprocessProgress}), more free neutrons remain (middle panel of Fig. \ref{fig:rprocessProgress}), 
and a robust main $r$ process results (bottom panel of Fig. \ref{fig:rprocessAbundance}).

If there is appreciable mu/tau emission from the disk, then the electron capture rates will 
not drop so steeply in the MNR region and may even increase, as described in Sec. 
\ref{sec:results} and shown in Figs. \ref{fig:nex024e2s20b2.0Multi20Percent} and 
\ref{fig:nex024e2s20b2.0Multi65Percent}. To explore this effect on the element synthesis we 
repeat the oscillation and outflow calculations described above for increasing percentages. 
Since these are multiple disk 
calculations, we adjust the radius of the emission surface of the mu/tau neutrinos 
 so that it corresponds to 10\%, 20\%, and 65\% of mu/tau emission as compared with 
electron antineutrino emission.  The results are shown in Figs. \ref{fig:rprocessAbundance} 
and \ref{fig:rprocessProgress} by the green, yellow, and red lines, respectively. The 
symmetric MNR influences both the electron neutrino and antineutrino fluxes, effectively 
swapping them with hotter but weaker mu/tau fluxes. Thus for a short time before alphas form, 
the balance in the forward reaction rates of Eq. \ref{eqn:weak} is adjusted by the MNR and the 
balance of protons and neutrons correspondingly shifts as shown in the middle panel of Fig. 
\ref{fig:rprocessProgress}. In the 10\% case, the neutrino capture rates still decrease, but 
there are enough electron neutrinos after the transition that a modest alpha effect occurs. 
This means that there are not enough neutrons for the r-process to get all the way to the 
highest mass number nuclei, although the nucleosynthesis does proceed beyond the $A\sim 80$ 
$r$-process peak region and some $A\sim 130$ peak nuclei are produced. In the 20\% case, the 
neutrino capture rates are relatively unchanged but the antineutrino capture rates are 
reduced, so that the overall composition is more proton rich than in the no oscillation case.  
This combined with a robust alpha effect results in a sharp reduction in $A\sim 80$ 
nucleosynthesis compared to the no oscillation case. With 65\% mu/tau emission, the MNR 
results in \emph{faster} neutrino capture rates, and the composition shifts to roughly equal 
numbers of neutrons and protons before alphas begin to form. Here only iron peak nuclei 
result.

\section{Conclusions}\label{sec:conclusion}

Compact object mergers emit large numbers of neutrinos with a flux that is initially composed 
of more electron antineutrinos than neutrinos.  This makes this environment a prime candidate 
for matter neutrino resonance transitions.  There are two types of relevant neutrino emission 
configurations: one where all flavors of neutrinos are emitted from the same surface, and one 
where neutrinos are emitted from different size surfaces.  The latter is more in line with the 
results of recent compact object merger simulations.  We find that two types of matter 
neutrino resonances occur. Both configurations show a standard MNR, which produces a 
transition where electron neutrinos wind up transformed into other flavors and electron 
antineutrinos wind up back in their initial eigenstates. We also find a symmetric MNR, where 
both electron neutrinos and antineutrinos are nearly completely transformed.  The type of 
transition is a consequence of the behavior of the neutrino self-interaction potential which 
depends sensitively on the balance of $\nu_e$ and $\bar{\nu}_e$s.

The size of the initial contribution of mu and tau type neutrinos to the flux, as compared 
with electron antineutrinos and neutrinos, is crucial to determining whether a transition 
occurs or not.  Comparable contributions will shut off the transition, but such large 
$\nu_\mu$ and $\nu_\tau$ fluxes are not currently predicted.

Future calculations of neutrino transport in mergers will be instrumental in determining what 
oscillation pattern we can expect above real world mergers, both because it is the balance of 
$\nu_e$ and $\bar{\nu}_e$s that determines the initial potential, and also because the mu and 
tau fluxes determine the flexibility available to the system for a MNR transition.

Matter-neutrino resonance transitions have a significant impact on wind nucleosynthesis, 
because they occur when the neutrino self-interaction potential and matter potential are 
approximately balanced.  If such transitions occur, they often occur close to the emission 
surface of the neutrinos, where the neutrinos still have large enough flux to affect the 
neutron to proton ratio, and nuclei in the outflow have not yet begun to form.

\begin{acknowledgments}
  This work was supported in part by U.S. DOE Grants No. DE-FG02-02ER41216 (GCM), DE-SC0004786 (GCM),
and DE-SC0013039 (RS).
\end{acknowledgments}
\bibliography{merger}

\end{document}